\documentclass[twocolumn,prb,amsmath,amssymb,amsfonts]{revtex4}
\usepackage{graphicx}

\begin{document}

\title{A widely tunable laser frequency offset lock with digital counting}

\author{Joshua Hughes}
\affiliation{Department of Physics and Astronomy, University of Georgia,
Athens, Georgia 30602\\}

\author{Chad Fertig}\thanks{e-mail: cfertig@uga.edu}
\affiliation{Department of Physics and Astronomy, University of Georgia,
Athens, Georgia 30602\\}

\date{\today}

\begin{abstract}
We demonstrate a hybrid analog+digital electronic lock to stabilize a dynamically tunable RF frequency offset between two lasers.  Our method features an 80~MHz capture range, $\pm$7~GHz tuning range, frequency agility of 1~MHz/$\mu$s, and low ($<30$~ppm) drift in the absolute optical frequency difference after $\sim$1000~s.  With this scheme, multiple slave lasers can easily be referenced to one stable master laser, while each remains rapidly and accurately tunable over the wide frequency ranges encountered in typical laser cooling and trapping experiments.
\end{abstract}

\maketitle
\section{Introduction}
Atomic laser cooling and trapping experiments typically require several independently tuned laser beams distributed in frequency across atomic ground and excited state manifolds. For example, in $^{87}$Rb the 5S$_\textrm{1/2}$ ground state and 5P$_\textrm{3/2}$ excited state hyperfine manifolds (connected by the 780~nm D2 line) have widths of 6.8~GHz and 500~MHz, respectively. In the course of a typical experiment it is often necessary to rapidly and accurately scan or jump the frequency of one or more lasers over these ranges, while maintaining absolute frequency reference to a particular transition.

To meet these goals, we have developed a scheme to electronically lock a rapidly, widely and accurately tunable frequency difference between two lasers based on digitally counting their optical beat frequency.  Here we report on two particular implementations of the scheme: a ``small-$\Delta$'' lock, in which a slave laser is locked 120~MHz above a stable master laser; and a ``large-$\Delta$'' lock, in which the slave is locked 6.7~GHz above the master, using nearly the same electronics.

\section{Structure of the paper}
The remainder of the paper is structured as follows.  In Section \ref{sec:compare} we review a number of published alternative schemes for locking the frequency difference between two lasers. In Section \ref{sec:description} we present a general description of our method, specializing to the two configurations realized in our laboratory (viz., the small- and large-$\Delta$ locks described above). In Section \ref{sec:signaltonoise} we analyze theoretically the performance limits set by quantization error for a frequency servo loop based on sampled digital counting. In Section \ref{sec:experiment} we present the results of experiments assessing the stability, accuracy and agility of the locks.  We conclude in Section \ref{sec:conclusion} with a brief discussion of potential improvements.

\section{Comparison to other schemes}
\label{sec:compare}
A large number of techniques to stabilize the frequency difference between two lasers have been developed over the long history of laser spectroscopy.  Here we review two broad categories that are most closely related to our method, and (not coincidentally) are often employed in laser cooling and trapping laboratories. 

In \emph{optical injection locks} a weak ``seed'' beam derived from a master laser is launched into the cavity of a slave laser, forcing the slave to lase on a longitudinal mode having frequency equal to that of the seed.  Frequency offsets between master and slave of 100's of MHz can be achieved by frequency shifting the seed beam with an acousto-optic modulator (AOM). \cite{Bouyer1996}  In this case, not only is the frequency difference between the lasers locked, but their optical fields are in fact phase coherent. Disadvantages are that careful optical mode matching of the seed beam to the slave laser is required; that the dynamical tuning range is limited by the modulation bandwidth of the AOM (typically less than 50~MHz); and that the range of possible offset frequencies is limited by the small selection of commercially available AOMs.

Replacing the AOM with an electro-optic modulator\cite{Kasevich1992} (EOM) can extend the dynamical tuning range beyond 100~MHz. EOMs are available over a wide range of operating frequencies up to 10's of GHz. However, these benefits come at considerably higher cost and additional complexity, to the extent that optical injection locking with EOM modulation is typically used only when phase-coherence is essential. 

It is also possible to use FM sidebands imposed on the master laser as seed beams.\cite{Cummings2002,Boyer2004}  While a large range of offsets can be achieved with this technique, it is only suitable for diode master lasers, and when only one or possibly two slaves are to be locked to one master, as it is challenging to isolate the effects of the spatially overlapping and co-propagating sidebands.

\emph{Electronic offset locks}, the family to which ours belongs, work by electronically detecting the optical beat note between beams from the master and slave lasers superimposed on a fast photodiode, then processing this signal to steer the frequency of the slave laser to maintain the desired offset.  An electronic offset lock designed around a commercial frequency-to-voltage (F-to-V) IC is described in Ref.~[4].  Unfortunately, commercial F-to-V chips are limited in operating frequency to at most 1~MHz, necessitating a deleteriously large prescaling factor to realize our small-$\Delta$ lock, and rendering unfeasible our large-$\Delta$ lock.  Furthermore, the charge pump output stage of commercial F-to-V chips typically exhibits a large (20\%) ripple that leads to undesirable dithering of the slave laser frequency even with aggressive low-pass post-filtering.  The hybrid analog+digital locking scheme we demonstrate here realizes high performance F-to-V conversion by using fast AC CMOS electronics to digitally sample and count the beat note frequency and a 10-bit DAC to generate an analog error signal suitable for processing by analog servo electronics.

Ref.~[5] describes an electronic offset lock based on a RF Mach-Zehnder interferometer. The RF beat note is split, transmitted through two unequal lengths of coaxial cables and recombined in an RF power combiner. The length of one cable is manually adjusted to produce, through total destructive interference, a node in the total transmitted power when the difference frequency between the two lasers is the desired value.  This scheme has the advantages of large capture range and a continuous selection of possible offset frequencies.  Unfortunately, since the offset frequency is tuned by a manual adjustment of the cable length, real-time, rapid tuning of the offset is impossible without additional frequency shifting components (e.g., AOMs).  Another serious drawback is the lack of absolute frequency reference for the offset, which is determined by the physical dimensions of cables.

Finally, electronic optical phase-locked-loops (OPLL) \cite{Zhu1993,Santarelli1994,Ricci1995,Muller2006} have been used to lock two lasers with a tunable offset.   Our lock is well suited to experiments that do not warrant the complexity of an OPLL, or that involve broad-line lasers for which the wideband control demands of on OPLL would be difficult to meet.  (Our control loop has a bandwidth of $\sim100$~kHz, but there is no critical level that must be met for stable operation, unlike for an OPLL.)  Also, our scheme allows for multi-GHz tunable offsets without the need for a tunable local oscillator in the microwave range, as is typically employed in OPLLs.

Our hybrid analog+digital electronic frequency offset lock is appropriate when a single, low-cost, flexible solution is required for multiple locks spanning a wide range of offsets from one stable master laser.

\section{Description of the method}
\label{sec:description}
\begin{figure*}
\includegraphics*[]{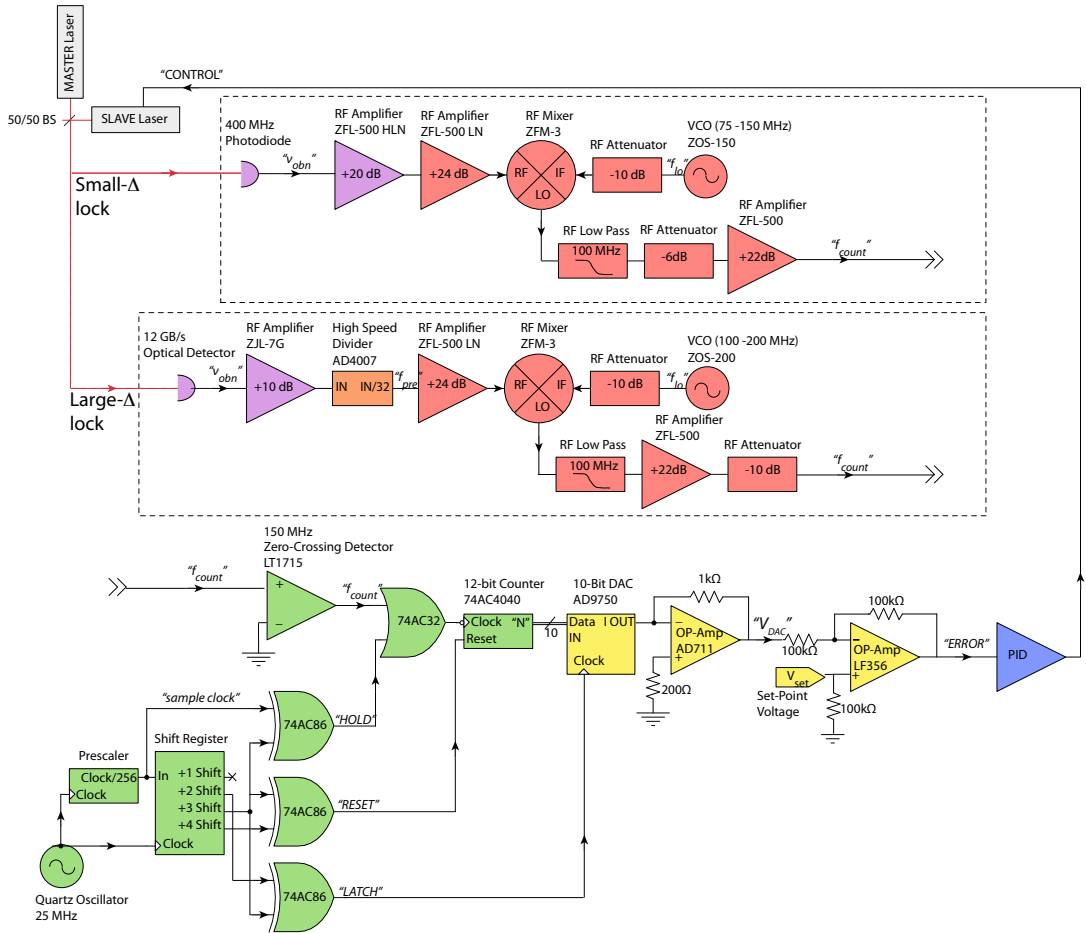}
\caption{\label{fig:blockdiagram}(Color online) Block diagram of circuit. Red arrows represent laser light, black arrows electronic signals.  Violet symbols: detection and amplification of the optical beat note. Red symbols: heterodyning stage. Orange symbols: fast prescaling stage for large-$\Delta$ lock. Green symbols: digital counting. Yellow symbols: D-to-A and generation of error signal. Blue symbols: PID feedback control.}
\end{figure*}

Fig. \ref{fig:blockdiagram} shows the essential optical and electronic components of our scheme.  A few milliwatts of light from the master laser (optical frequency $\nu_m$) and slave laser (optical frequency $\nu_s$) are combined on a fast photodetector. The induced photocurrent comprises a DC component proportional to the total incident optical power and an AC component oscillating at the optical beat note frequency $\Delta\nu_{\textrm{obn}} = \nu_m - \nu_s$.  The AC signal is amplified, prescaled in the case of the large-$\Delta$ lock, shifted via heterodyne mixing, and ultimately counted with digital electronics.  The digital frequency counting starts with a fast zero-crossing detector generating a train of logic pulses synchronous with its analog input.  The number of logic pulses (typically 200) in each sample window of duration $\mathcal{T}\approx5~\mu$s is counted by a gated 12-bit binary counter.  At the conclusion of each sample window the counter's data is latched into a digital-to-analog converter (DAC) that generates an output voltage $V_{\textrm{DAC}}$ proportional to the average frequency difference between the two lasers during that window.  Subtracting $V_{\textrm{DAC}}$ from a computer-controlled set-point voltage $V_{\textrm{set}}$ in a differencing amplifier produces an error signal suitable for subsequent processing by a PID (proportional-integral-differential) servo controller.  The servo controller generates an electronic control signal that steers the slave laser frequency to maintain the desired offset with respect to the master.

We implement the digital electronics using AC CMOS logic, which has a theoretical maximum operating frequency of $\sim120$~MHz. Since both our small- and large-$\Delta$ locks are intended to maintain values of $\Delta\nu_{\textrm{obn}}$ at or exceeding this limit, we must take steps to translate the frequency of the optical beat note down into the working range of our counting electronics. For example, to count the $\Delta\nu_\textrm{obn}=120$~MHz beat note of our small-$\Delta$ lock, we heterodyne the optical beat note with a local oscillator of frequency $f_\textrm{LO}=80$~MHz in an RF mixer, then digitally count the \emph{down}-shifted copy of the beat note at $f_\textrm{count}=\Delta\nu_{\textrm{obn}} - f_{LO}=40$~MHz.  Besides shifting the beat note into the working range of the counting electronics, the heterodyning step has other significant benefits.  First, it provides an attractive method to tune the master-slave frequency difference by adjusting $f_{LO}$ while leaving the reference voltage $V_\textrm{set}$ centered in the circuit's capture range.\cite{footnote1} Second, by counting the \emph{up}-shifted copy of the beat note from the RF mixer, we can lock a master-slave offset frequency for which the optical beat note would otherwise be too low in frequency to digitally count with the fidelity required for a high-bandwidth servo loop.

To count the $\Delta\nu_\textrm{obn}=6.7$~GHz beat note of our large-$\Delta$ lock we use a high bandwidth photoreceiver (Discovery Semiconductor DSC-R402-89) and divide the frequency of the optical beat note by 32 using a high frequency divider IC, producing a signal at the prescaled frequency $f_{\textrm{pre}}=\Delta\nu_{\textrm{obn}}/32=209$~MHz. The prescaled signal is then RF-heterodyned as in the small-$\Delta$ lock, though with $f_{LO}=174$~MHz, to produce a (down-shifted) signal at $f_\textrm{count}=f_\textrm{pre}-f_\textrm{LO}=35$~MHz, suitable for digital counting.

\begin{figure*}
\includegraphics*[]{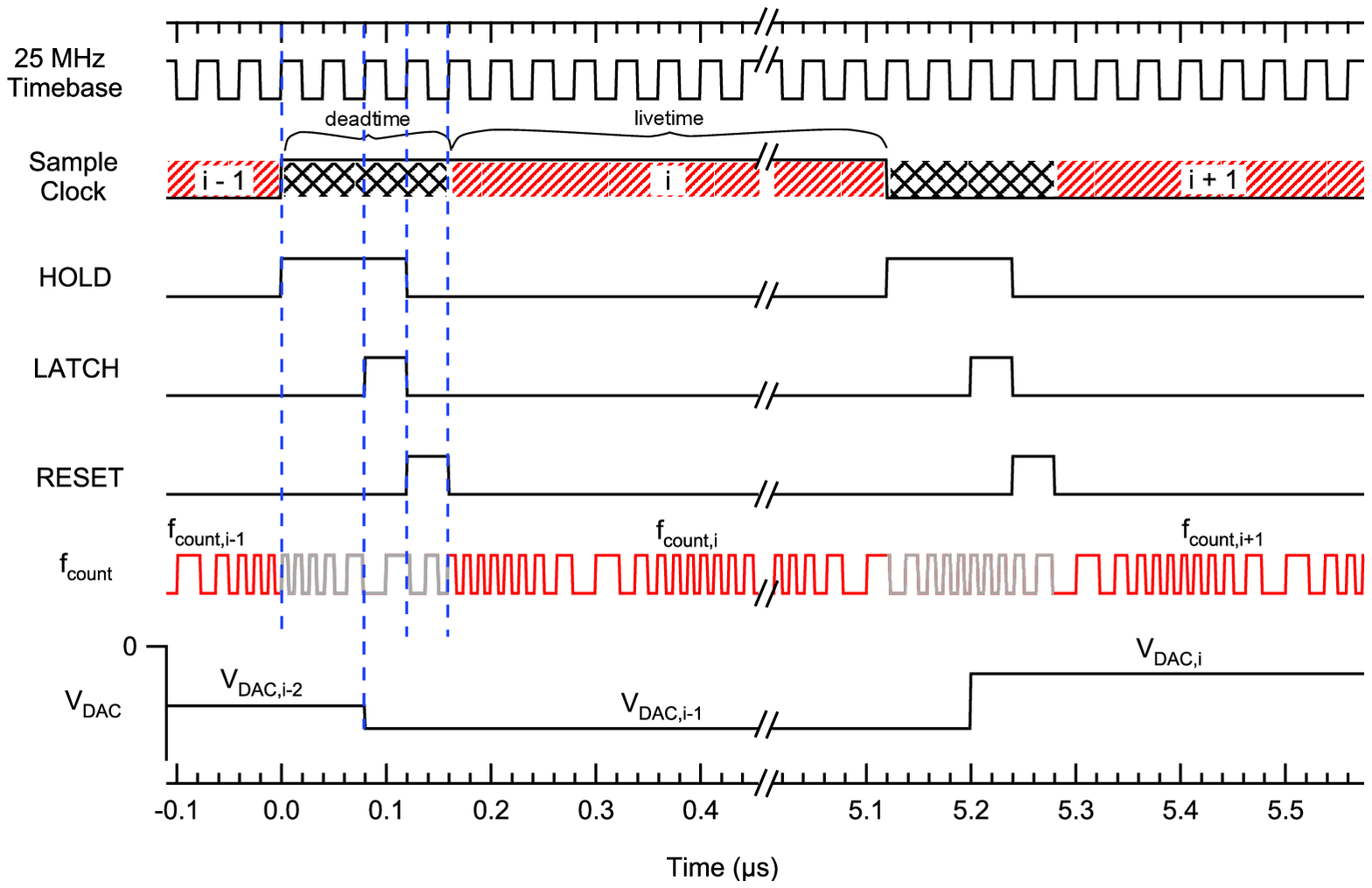}
\caption{\label{fig:timingdiagram} (Color online.) Timing diagram of digital counting.  Vertical blue dashed lines are guides to the eye.  Counting occurs only in red shaded regions of the sample clock waveform, yielding a livetime fraction of $\sim$97\% (note the broken time axis).  $V_\textrm{DAC}$ is updated at the conclusion of a given counting window; for example, $V_\textrm{DAC,i-1}$, proportional to the average counted frequency in the $(i-1)$ sample window, is output by the DAC while the binary counter is accumulating counts in the $i$-th window.  This $\sim5~\mu$s latency adversely effects the stability of the servo loop at frequencies near the Nyquist frequency $1/\mathcal{T}\approx 200$~kHz, and thus imposes a limit on the loop gain.}
\end{figure*}

\label{subsec:digital}
The post-heterodyne signal at $f_\textrm{count}$ is input to a high speed zero-crossing detector that generates a synchronous AC CMOS digital signal.\cite{footnote2}  This signal is binned and counted in sample windows of duration $\mathcal{T}=4.96~\mu$s at a rate of 195.3~kSamples/s. The sample clock is derived from a 25.000~MHz crystal oscillator divided by 256 with an 8-bit binary counter.  As shown in Figs.~\ref{fig:blockdiagram} and \ref{fig:timingdiagram}, a multiple-output shift register generates three auxiliary logic signals from the 195.3~kHz sample clock, shifted by +2, +3, and +4 cycles of the 25 MHz crystal oscillator (i.e., by 80, 120 and 160~ns, respectively).  The unshifted and ``+3'' signals are XOR'd to generate the ``HOLD'' signal; ``+2'' and ``+3'' are XOR'd to generate the ``LATCH'' signal; and ``+3'' and ``+4'' are XOR'd to generate the ``RESET'' signal. The rising edge of RESET asynchronously clears the 12-bit counter; counting commences 40~ns later, after the fall of RESET, and continues for $\mathcal{T}=4.96~\mu$s (the ``sample window''), until HOLD toggles high. (A logical-high HOLD suspends counting by preventing falling edges from the zero-crossing detector from arriving at the 12-bit counter.) As long as RESET and HOLD are both low, the 12-bit counter will count logic pulses from the zero-crossing detector.  The 80~ns delay between the close of the counting window and the start of data conversion by the DAC (on the rising edge of LATCH) is included to satisfy a minimum data setup time specification of our DAC. The result of this gating logic is a livetime fraction of approximately 97\%.

The lowest 10 bits of the 12-bit counter serve as the inputs to a 10-bit DAC.  The DAC is operated in buffered, single-ended output configuration with an output voltage range of $V_\textrm{DAC} = 0~\textrm{to~} -11.6$~V. To generate the ``ERROR'' signal for the analog PID servo loop, $V_\textrm{DAC}$ is subtracted from the computer generated set-point voltage $V_\textrm{set}$ in a differencing amplifier.  We choose $V_\textrm{set}\approx-2\textrm{V to}~-3\textrm{V}$ separately for the small- and large-$\Delta$ configuration so that $f_\textrm{count}$ is approximately centered in the capture range when the servo loop is locked.

\section{Quantization error limited stability of a digitally counted frequency servo loop}
\label{sec:signaltonoise}

\begin{figure}
\includegraphics*[]{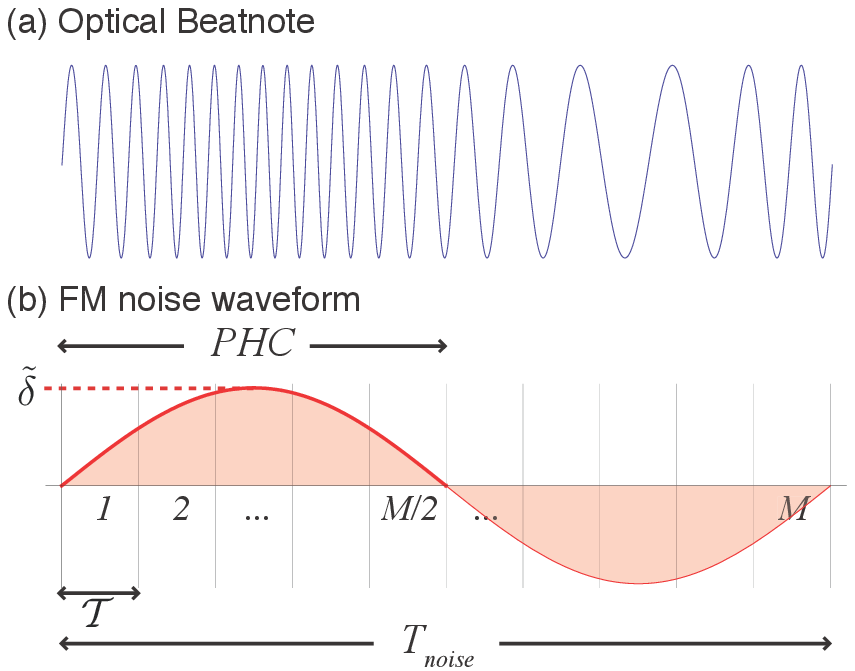}
\caption{\label{fig:signaltonoise} (Color online.) Detecting frequency fluctuations by discretely sampled counting.  (a) Pictorial representation of the optical beat note frequency modulated by the noise waveform of (b).  (b) Representative FM noise waveform, identifying various symbols used in the text.}
\end{figure}

There is a fundamental limit to the frequency stability of a lock such as ours that employs sampled digital frequency counting. Suppose that $N=M\mathcal{T}f$ counts (i.e., zero crossings) of a noiseless carrier at frequency $f$ are detected in $M$ sample windows of duration $\mathcal{T}$ (neglecting deadtime). Introduce monochromatic FM noise $\delta f(t)=\tilde{\delta} \sin(2 \pi f_\textrm{noise} t)$ on the carrier.  For simplicity we restrict our calculation to noise frequencies with period an integer number of sample windows: $T_\textrm{noise} = M \mathcal{T}, M=2,3,4,\dots$.  This FM noise on the carrier causes \emph{additional} counts\cite{footnote3} $$\Delta=\int_\textrm{\tiny PHC} \delta f (t)~dt=\frac{T_\textrm{noise}\tilde{\delta}}{\pi}$$
 during the noise waveform's positive half-cycle (``NWPHC,'' from $t=0$ to $t=T_\textrm{noise}/2$ (see Fig. \ref{fig:signaltonoise}), where $T_\textrm{noise}=1/f_\textrm{noise}$.  We define the \emph{minimum detectable noise amplitude} $\tilde{\delta}=\tilde{\delta}_\textrm{min}$ as that fluctuation of the carrier frequency that causes exactly $\Delta=1$ to accumulate over the NWPHC. With the perspective that the ``signal'' we wish our circuit to detect (and subsequently to correct via feedback) is the instantaneous frequency fluctuation $\delta f(t)$, and that the ``noise'' that disturbs this effort is the unavoidable $\pm 0.5$ count uncertainty in the least significant bit of the binary counter, we say that the fluctuation $\tilde{\delta}_\textrm{min}$ will be sensed by the circuit with a quantization error limited (``QEL'') signal-to-noise ratio $S/N \sim 1,$ neglecting all other sources of noise.  A fluctuation smaller than $\tilde{\delta}_\textrm{min}$ would go unnoticed (and therefore uncorrected) by the circuit, while larger fluctuations would be sensed (and corrected) with $S/N > 1$. 

Using relations developed in the preceding discussion, we find $$\tilde{\delta}_\textrm{min}=\frac{\pi}{T_\textrm{noise}}.$$  In a sampled digital frequency counter such as ours, the minimum detectable frequency fluctuation of the carrier decreases with increasing FM noise period (decreasing FM noise frequency). This imposes a fundamental stability limit to a carrier whose frequency is controlled by a feedback loop that employs such a counter. In our lock, the ``carrier'' is the optical beat note at $\Delta\nu_\textrm{obn}$. The QEL implies that fluctuations of the average optical beat note frequency between successive time-averages of duration $\tau$ cannot be smaller than $\tilde{\delta}_\textrm{min}(\tau)=\pi/\tau,$ and thus will exhibit a (QEL) relative rms deviation of no less than $$\sigma^\textrm{QEL}_\textrm{y}(\tau)=\frac{\pi}{\sqrt{2}\tau\Delta\nu_\textrm{obn}}.$$

The division of the optical beat note by 32 in the large-$\Delta$ lock increases $\tilde{\delta}_\textrm{min},$ and therefore the \emph{absolute} frequency instability, by the same factor.  However, since the large-$\Delta$ lock prescaled frequency $f_{\textrm{pre}}=209$~MHz is $1.7\times$ larger than the unscaled optical beat note $\Delta\nu_{\textrm{obn}}=120$~MHz of the small-$\Delta$ lock, the QEL \emph{relative} instability limit is only a factor of $32/1.7\approx19$ higher (i.e., worse) for the large-$\Delta$ lock.  Figs.~\ref{fig:Allan} shows lines demarcating the QEL instability limits for both locks, in both relative and absolute units.

\section{Experimental characterization}
\label{sec:experiment}
Here we report the results of several experiments to characterize the performance of both small- and large-$\Delta$ locks. To facilitate comparison of the two configurations, the same master laser, slave laser and lock electronics were used in all experiments, save for the explicit differences indicated in Fig. \ref{fig:blockdiagram}.	

\begin{figure}
\includegraphics*[]{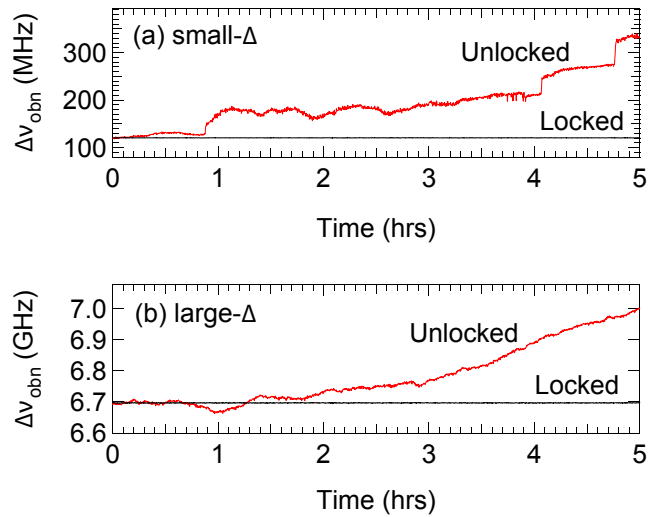}
\caption{\label{fig:longterm}(Color online.) Long-term locking.  Time series of the optical beat note frequency $\Delta\nu_{\textrm{obn}}$, acquired over a 5 hour period, for the small-$\Delta$ (a) and large-$\Delta$ locks (b). For comparison, time series are also shown for the slave laser unlocked.}
\end{figure}

\begin{figure}
\includegraphics*[]{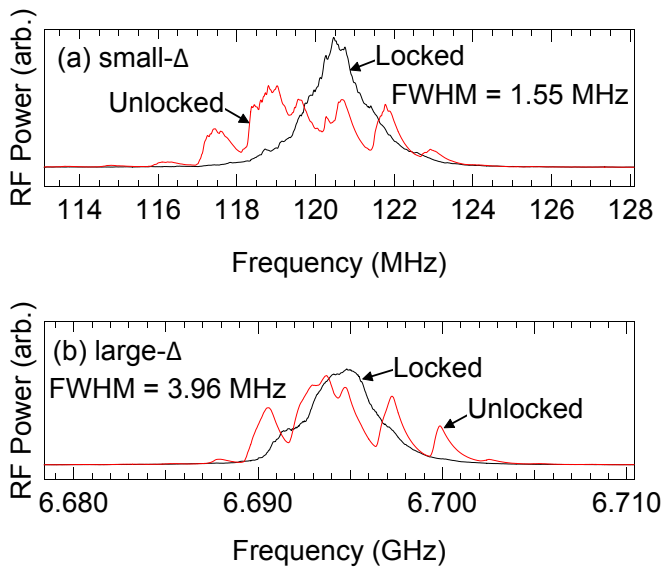}
\caption{\label{fig:noise}(Color online.) Optical beat note linewidth.  RF spectrum analyzer traces of the optical beat note for small-$\Delta$ (a) and large-$\Delta$ (b) locks for the servo loops both locked and unlocked.  FWHM linewidths are given for the locked data.  For these measurements, the resolution bandwidth of the spectrum analyzer was 10~kHz (100~kHz) for the small- (large-)$\Delta$ locks, respectively.  The full-screen sweep time was approximately 20~ms for all traces.}
\end{figure}

The master and slave lasers are both grating stabilized, external-cavity diode lasers with $\sim~1$~MHz intrinsic ``fast'' linewidths. Fig. \ref{fig:longterm} shows a 5~hr time series of the optical beat note frequency for both locks under continuous operation, measured at 10~s intervals with an RF spectrum analyzer.

We evaluated the short time stability of the optical beat note by directly measuring its linewidth over a 10~ms integration time using an RF spectrum analyzer (see Fig. \ref{fig:noise}).  The beat note linewidth was measured to have FWHM of 1.5~MHz (4.0~MHz) for the small- (large-)$\Delta$ locks, respectively.  The broadening of the large-$\Delta$ linewidth is qualitatively consistent with quantization error limited performance (see Sec.~\ref{sec:signaltonoise}).

\begin{figure}
\includegraphics*[]{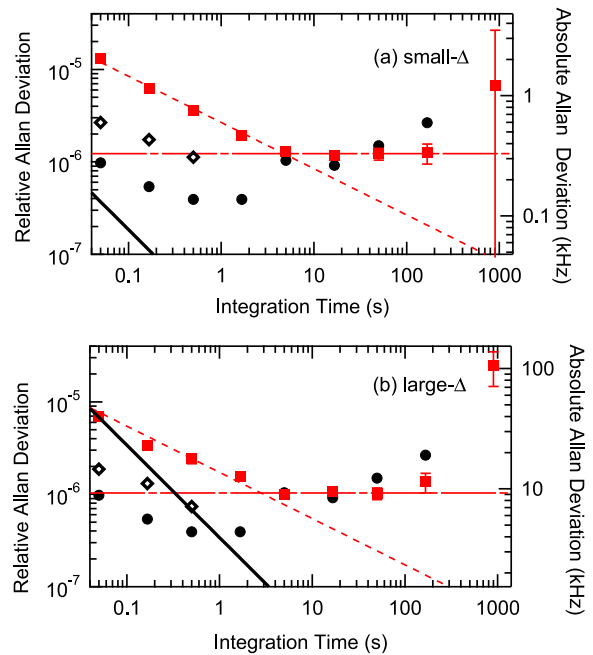}
\caption{\label{fig:Allan} (Color online.) Frequency stability. Absolute and relative frequency Allan deviation data are shown as red solid squares for the small-$\Delta$ (a) and large-$\Delta$ (b) locks, for integrations times $0.05~\textrm{s}\leq\tau\leq167~\textrm{s}$..  Points at 900~s are not true Allan deviations but rather the average deviation of 13 discrete measurements of the beat note taken at 15 minute intervals over 3 hours. The dashed and dashed-dot lines are fits to the measured Allan deviation points, as described in the text.  Heavy black lines in both figures show the quantization error limited (QEL) relative instability $\sigma^\textrm{QEL}_\textrm{y}(\tau)$ discussed in Sec.~\ref{sec:signaltonoise}.  Solid black circles are the measured relative Allan deviation of the computer-controlled reference voltage.  Open black diamonds show the counting noise floor of the counting electronics as measured in a separate experiment in which the optical beat note was replaced by the output of a high-quality electronic synthesizer set to the same frequency, with the high speed divider either removed (a), or set to divide by 16 (b).  There is no contradiction in these measurements lying below the QEL line, as they were made in open loop, whereas the QEL applies only to the closed loop stability of the \emph{locked} laser beat note.  Error bars are 1-standard deviation statistical uncertainties.}
\end{figure}

To characterize frequency stability over longer times, a second, independent copy of the lock electronics was used to make an ``out of loop'' measurement of the frequency Allan deviation $\sigma_y(\tau)$ of the beat note for integration times $0.05~\textrm{s}\leq\tau\leq167~\textrm{s}$; these data are presented in Fig. \ref{fig:Allan}. To make these measurements, fluctuations of the error signal voltage of the out-of-loop counting box were recorded using a digital multimeter, and the measured voltage noise subsequently converted to frequency fluctuations.  The deadtime fraction for the measurements is $<12\%$ and is ignored in the analysis, except for the points at $\tau=900$~s (see Fig.~\ref{fig:Allan} caption).  For the large-$\Delta$ lock, the data is fit to a white-noise-limited model for $0.05~\textrm{s}\leq\tau\leq5~\textrm{s}$ with a result $\sigma_y(\tau)=1.6\times10^{-6}\tau^{-1/2}$.  A 1/f noise dominated ``flicker floor'' model is fit to the data for $5~\textrm{s}\leq\tau\leq167~\textrm{s}$ with the result $\sigma_y(\tau)=1.0\times10^{-6}$.  Similar fits are made to the small-$\Delta$ lock data, yielding $\sigma_y(\tau)=2.6\times10^{-6}\tau^{-1/2}$ for $0.05~\textrm{s}\leq\tau\leq5~\textrm{s},$ and $\sigma_y(\tau)=1.2\times10^{-6}$ for $5~\textrm{s}\leq\tau\leq167~\textrm{s}$.  These fits are shown in Fig.~\ref{fig:Allan}.  We could not measure Allan deviations for integration times sufficiently short to probe the onset of QEL performance for either lock.  By extrapolating the data we estimate that the small- (large-)$\Delta$ locks would reach QEL at $\tau=5\times10^{-5}$~s ($\tau=0.045$~s), respectively.

Fig.~\ref{fig:Allan} also shows measurements of the Allan deviation of the reference voltage $V_\textrm{set}$, which in our system is supplied by a PC-based multifunction card, which are consistent with the observed level of the ``flicker floor'' for both locks. Inspection of the raw time-series data show that the increase in instability of $V_\textrm{set}$ at $\tau=1$~s is the result of rare, discrete jumps in $V_\textrm{set}$.  We do not have an explanation for the jumps, which are much smaller than can be attributed to LSB noise in the PC card's output DAC.  In the future we plan to replace the PC card with a higher-performance model, upgrade electrical connections and install opto-isolators to break ground loops, in an attempt to suppress these jumps.

\begin{figure}
\includegraphics*[]{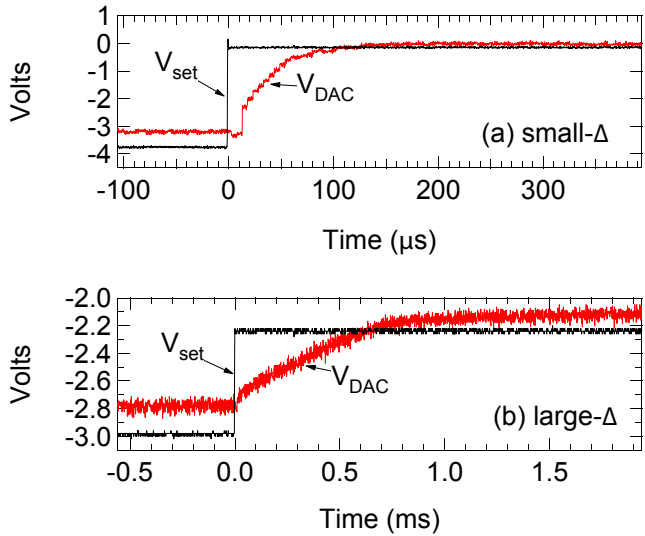}
\caption{\label{fig:jump}(Color online.) Tuning agility. Oscilloscope traces of $V_\textrm{DAC}$, proportional to $\Delta\nu_{\textrm{obn}}$, show the response of the slave laser to an abrupt step of $V_\textrm{set}$, corresponding to jumps of 70~MHz in the small-$\Delta$ lock (a) and 416~MHz in the large-$\Delta$ lock (b).  The stair-case pattern visible in $V_\textrm{DAC}$ in (a) directly reflects the discrete sampling and counting of the beat note. The steps are not visible in (b) due to its compressed time axis.}
\end{figure}

We quantify the tuning agility of the lock as the average frequency slewing rate over the time it takes the beat note to track from 10\% to 90\% of its asymptotic value (``10-90'' time), following a large step in $V_\textrm{set}$. Fig. \ref{fig:jump} shows traces of $V_\textrm{DAC}$, proportional to $\Delta\nu_{\textrm{obn}}$, from which we extract a 10-90 time of 71~$\mu$s (690~$\mu$s) after a 70~MHz (416~MHz) step, for an average slew rate of 1.0~MHz/$\mu$s (1.7~MHz/$\mu$s) for the small- (large-)$\Delta$ locks, respectively.  These results are consistent with the designed 100~kHz bandwidth of the servo loop.

\begin{figure}
\includegraphics*[]{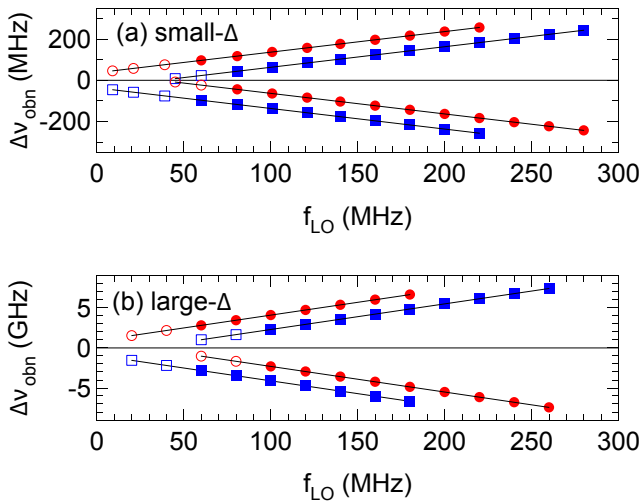}
\caption{\label{fig:accuracy}(Color online.) Tuning range. The frequency of the locked optical beat note $\Delta\nu_{\textrm{obn}}$, for the small-$\Delta$ lock (a) and large-$\Delta$ lock (b) versus heterodyne local oscillator frequency $f_\textrm{LO}$, counted using an RF spectrum analyzer.  A positive (negative) beat note corresponds to the slave laser frequency higher (lower) than the master laser. Solid red circles (solid blue squares) correspond to locking to the down- (up-)shifted copy of the beat note from the heterodyne stage.  Open symbols represent measurements made with a different value of the post-heterodyne RF low-pass filter so as to lock the lowest optical beat notes for a given configuration. Curves through the data are lines of slope=1 (=32) through the lowest measured beat note of each series, for the small- (large-)$\Delta$ locks, respectively.}
\end{figure}

The optical beat note can be accurately tuned over a wide range by adjusting the heterodyne local oscillator frequency $f_\textrm{LO}$.  Fig.~\ref{fig:accuracy} shows data demonstrating the range and accuracy of closed-loop tuning of the slave laser in which $f_\textrm{LO}$ is changed in 20~MHz steps for both the small- and large-$\Delta$ locks. (For the large-$\Delta$ lock, because of the division-by-32 prescaling of the optical beat note, the tuning steps are $32\times20\textrm{~MHz}=640$~MHz.) While the 80~MHz capture range of the servo loop limits the size of an individual abrupt step, smooth adjustments to $f_\textrm{LO}$ permit tuning limited on the high-end only by the bandwidth of the RF photodetector and on the low-end by the steepness of the post-heterodyne low-pass filter.\cite{footnote4}  As discussed in Section \ref{sec:description}, we reach the lowest lockable beat notes by selecting the up-shifted heterodyne component, while choosing the polarity of the servo loop appropriate to the sign of $\Delta\nu_\textrm{obn}$ so as to yield overall negative feedback.

\section{Conclusion}
\label{sec:conclusion}
We have designed and implemented a hybrid analog+digital scheme to lock the difference frequency between two lasers. The long term drift, short term stability, tuning range, tuning accuracy and overall flexibility of the scheme make it well suited to applications where the frequencies of several independent lasers need to be stabilized to an absolute reference \emph{and} rapidly and accurately tuned over several GHz.  Improvements to the digital electronics, including optimized component layout, better ground plane practices, and the use of stripline leads for signal transport should allow the present design to reach the maximum clock rate for AC CMOS logic of 120~MHz, for a modest increase in performance.  A large jump in performance could be realized by switching to a higher speed logic family, such as ECL, which would enable count rates of 500~MHz or more, improving the signal-to-noise, capture range, and closed-loop bandwidth of the lock. It would then be feasible to implement fully digital-domain processing of the error signal, which could greatly improve both the long-term stability and absolute accuracy of the lock.  

In conclusion, we have demonstrated a method to lock the offset frequency between two lasers based on the direct digital counting of the optical beat note, which is well suited to atomic laser cooling and trapping experiments that call for many widely tunable laser beams, each with absolute frequency reference.

This work was sponsored in part by grants from the ARO, the DURIP, the ORAU, and the University of Georgia.


\begin{thebibliography}{13}
\bibitem[{foo({\natexlab{a}})}]{footnote1}
\bibinfo{note}{While the upper end of the capture range is straightforwardly
  set by the maximum operating frequency of the digital electronics, the lower
  end is (less obviously) set by the frequency at which the master and slave
  lasers cross (i.e., $\Delta\nu_\textrm{obn}$ changes sign), or at which
  $f_\textrm{count}$ crosses $\Delta\nu_\textrm{obn}$, whichever is higher. It
  is at this point that the ``lock'' becomes an ``anti--lock'' (i.e., the servo
  loop feedback becomes positive, rather than negative).}
\bibitem[{foo({\natexlab{b}})}]{footnote2}
\bibinfo{note}{For reasons that are not fully understood, our zero-crossing
  detector was fairly sensitive to input level, which was fine-tuned to
  optimize performance.}
\bibitem[{foo({\natexlab{c}})}]{footnote3}
\bibinfo{note}{We neglect issues related to the relative phases of the counted
  signal, the noise waveform and the sample clock; such considerations affect
  the results of this analysis at the fractional level of $1/MN$, which are
  never greater than 1\% for us.}
\bibitem[{foo({\natexlab{d}})}]{footnote4}
\bibinfo{note}{The lock range is also (obviously) limited by the mode-hop free
  tuning range of the slave laser.}
\end{thebibliography}
\end{document}